\documentclass[article]{JHEP3}
\usepackage{graphicx}
\usepackage{multirow}
\usepackage{cite}

\newcommand\clH{{\mathbb H}}
\newcommand\clS{{\mathbb S}}
\preprint{CP3-14-61, MCnet-14-18, LPN14-105}

\title{Top-quark mass effects in double and triple Higgs production in gluon-gluon fusion at NLO}

\author{F.~Maltoni$^a$, E.~Vryonidou$^a$, M.~Zaro$^{b}$\\
$^a$ Centre for Cosmology, 
 Particle Physics and Phenomenology (CP3),\\
 Universit\'e Catholique de Louvain, B-1348 Louvain-la-Neuve, Belgium \\
 $^b$ Sorbonne Universit\'es, UPMC Univ. Paris 06, UMR 7589, LPTHE, F-75005, Paris, France and
CNRS, UMR 7589, LPTHE, F-75005, Paris, France}
\abstract
{The observation of double and triple scalar boson production at hadron colliders could provide key 
information on the Higgs self couplings and the potential. As for single Higgs production
the largest rates for multiple Higgs production come from gluon-gluon fusion processes mediated by a top-quark loop. However, at variance with single Higgs production, top-quark mass and width effects from the loops cannot be neglected. Computations including the exact top-quark mass dependence are only available at the leading order, and currently predictions at higher orders are obtained by means of approximations based on the Higgs-gluon effective field theory (HEFT).  In this work we present a reweighting technique that, starting from events obtained via the MC$@$NLO method in the HEFT, allows to exactly include the top-quark mass and width effects coming from one- and two-loop amplitudes. We describe our approach and apply it to double Higgs production at NLO in QCD, computing the needed one-loop amplitudes and using approximations for the unknown two-loop ones. The  results are compared to other approaches used in the literature, arguing that they provide more accurate predictions for distributions and for total rates as well. As a novel application of our procedure we present predictions at NLO in QCD for triple Higgs production at hadron colliders.}

\maketitle

\begin{document}

\section{Introduction}
\label{sec:intro}

Present LHC data already provide convincing evidence that the scalar particle observed at the LHC is the one predicted by the Brout-Englert-Higgs breaking mechanism~\cite{Englert:1964et,Higgs:1964pj} of the $SU(2)_L \times U(1)_Y$ symmetry as implemented in the Standard Model (SM)~\cite{Weinberg:1975gm}.
Here, the strength of the Higgs boson couplings is uniquely determined by the masses of the elementary particles, including the Higgs boson itself. The measured couplings to fermions and vector bosons  are found to agree within 10-20\% with the SM predictions~\cite{cms2013,atlas2013}. No direct information, however, has been collected so far on the Higgs self-couplings that appear in the potential:
\begin{equation}
V(H)=\frac{1}{2}m_H^2 H^2 + \lambda_{HHH} v H^3 + \frac{1}{4} \lambda_{HHHH} H^4.
\end{equation}
The values of the Higgs self-couplings $\lambda_{HHH}$ and $\lambda_{HHHH}$ are fixed in the SM by gauge invariance and renormalisability to $\lambda_{HHH}=\lambda_{HHHH}=m_H^2/2v^2$, {\it i.e.} fully determined by the mass of the Higgs boson and the Higgs field vacuum expectation value $v$. Direct information on the Higgs three-point and four-point interactions would therefore provide key information on the upper scale of validity of the SM when thought of as an effective theory itself, or on the possible existence of a richer scalar sector, featuring additional scalar fields, possibly in other representations.

In this context, multiple Higgs production plays a special role. At the lowest order, Higgs pair production is the simplest production process that is sensitive to the trilinear self-coupling $\lambda_{HHH}$,
 while to probe the quartic Higgs coupling $\lambda_{HHHH}$ one would need  to consider at least
 triple Higgs production. Unfortunately, in the SM multiple Higgs production 
rates at the LHC are quite small~\cite{Plehn:1996wb,Plehn:2005nk} and
the prospects of making precise enough measurements at the LHC (assuming SM values) are 
at best challenging~\cite{Baur:2002rb,Baur:2003gpa,Baur:2003gp}
for double Higgs production and rather bleak for triple Higgs production~\cite{Plehn:2005nk,Binoth:2006ym}.

However, multiple Higgs production rates could be enhanced by new physics effects and therefore the process provides one with a wealth of possibilities for probing new physics.
These possibilities range from explicit models featuring new particles, 
such as the 2HDM \cite{Baglio:2014nea,Hespel:2014sla,Asakawa:2010xj,Arhrib:2009hc},
SUSY \cite{Plehn:1996wb,Cao:2013si,Ellwanger:2013ova,Nhung:2013lpa}, 
extended colored sectors~\cite{Dawson:2012mk,Kribs:2012kz,Chen:2014xwa},
Little Higgs Models \cite{Dib:2005re,Wang:2007zx}, Higgs portal~\cite{Dolan:2012ac,No:2013wsa},
flavor symmetry models \cite{Berger:2014gga}
and Composite Higgs models~\cite{Grober:2010yv,Gillioz:2012se}
to model independent higher-dimensional interactions~\cite{Contino:2012xk,Pierce:2006dh,Liu:2013woa}.
In any case, precise predictions for rates and distributions will be needed to be able to
  extract valuable information on $\lambda_{HHH}$ or on new physics effects in general. 
 Higgs pair production will be relevant for the high-luminosity LHC
 and several phenomenological studies
 \cite{Papaefstathiou:2012qe,Baglio:2012np,Dolan:2012rv,Barr:2013tda,Gouzevitch:2013qca,Li:2013flc,Goertz:2013kp,deLima:2014dta,Barger:2013jfa,Slawinska:2014vpa} have recently investigated the potential of detecting the signal for the process over the backgrounds in various Higgs decay channels. Triple Higgs production rates are smaller by more than two orders of magnitude~\cite{Plehn:2005nk,Binoth:2006ym}, 
 making the process possibly relevant for a future 100~TeV collider. 

Analogously to single Higgs production, several  channels can lead to a final
 state involving two or three Higgs bosons. They involve either the Higgs coupling
to the top quark ({\it e.g.},  gluon-gluon fusion via top loops and $t \bar t$ associated production), 
or to vector bosons ({\it e.g.},  vector boson fusion and associated production), 
or to both ({\it e.g.}, single-top associated production).
The dominant production mechanism for multiple Higgs  production is gluon-gluon fusion
 via top-quark loops, exactly as in the case of single Higgs production. 
 Cross sections corresponding to the other channels are at least one order of magnitude smaller, 
 even though possibly interesting because of different sensitivity to the $\lambda$'s or to other couplings,
 such as $VVHH$ \cite{Contino:2013gna,Pappadopulo:2014qza}, or to new physics.  In addition, typically, different channels open different possibilities of
  exploiting a wider range of Higgs decay signatures.
Predictions for the six main $HH$ production channels at NLO in QCD for hadron collider energies ranging
 from 8 to 100 TeV can be found in \cite{Frederix:2014hta}.  NNLO results for vector boson fusion \cite{Liu-Sheng:2014gxa} and vector boson associated production
  \cite{Baglio:2012np} are also available. For triple Higgs production
 gluon-gluon fusion is again the dominant production channel by at least one order of magnitude. However,  
 the importance of  the subdominant production channels is slightly different from the pair production 
 at 14 TeV, with $t \bar t$ associated production giving the second largest contribution,  and VBF the third.

Gluon-gluon fusion being the dominant production mechanism for multiple Higgs-boson production, one would like to have the most accurate and precise predictions possible
for this channel. However, as Higgs pair production is a loop induced $2 \to 2$ process at the Born level, higher-order computations become very involved.  Computations including the exact top-quark mass dependence, {\it  i.e.}  in the Full Theory (FT), are only available at the leading order, and currently predictions at NLO~\cite{Dawson:1998py} and NNLO accuracy~\cite{deFlorian:2013jea,Grigo:2014jma}  are obtained by means of approximations that build upon the Higgs-gluon effective field theory.
 Besides, at variance with the inclusive cross section computation for single Higgs production, top-quark mass effects in the loops cannot be neglected.
In this work we present a method that, starting from an NLO calculation matched to parton shower (PS) in the Higgs effective-field theory allows us to
systematically include the heavy-quark mass effects coming from the one- and two-loop amplitudes, as long as results for the latter are available. We  
describe our approach, validate it against single Higgs production, for which all the necessary loop amplitudes are known, and apply it to double 
and triple Higgs production using all available information, {\it  i.e.} the exact one-loop (real) amplitudes and the known parts of the two-loop amplitudes.
We dub our results  FT$_{\rm approx}$ at NLO(+PS) and compare them to other approximations in the literature. We show that with just 
the currently available information on the loop amplitudes, our reweighting method already provides more accurate results for the differential distributions
 at NLO plus parton shower level and argue that that is the case for total rates as well. 
 
The paper is structured as follows. In section 2 we discuss the current status of the computation for multiple Higgs production. In section 3 we describe our method
 and in particular how the top-quark mass effects can be consistently included in the calculation. In section 4 we present our results for Higgs pair production, comparing our method
  with previous results in the literature. In section 5 for the first time we present results for triple Higgs production beyond the leading order, obtained using the same method. 
 We summarise our findings and draw some conclusions in the final section.

\section{Multiple Higgs production in gluon-gluon fusion}

\begin{figure*}
 \center 
 \includegraphics[scale=0.6]{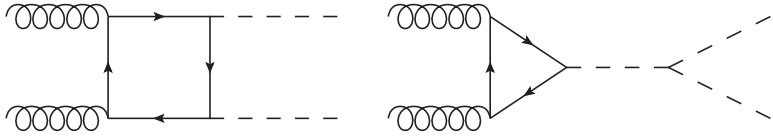}
 \caption{\label{fig:LO}
 Representative Feynman diagrams for box and triangle topologies for Higgs pair production in gluon-gluon fusion at the lowest order in perturbative QCD. The two gauge-indepedent classes of diagrams interfere destructively.  }
\end{figure*}

\begin{figure*}
 \center 
 \includegraphics[scale=0.86]{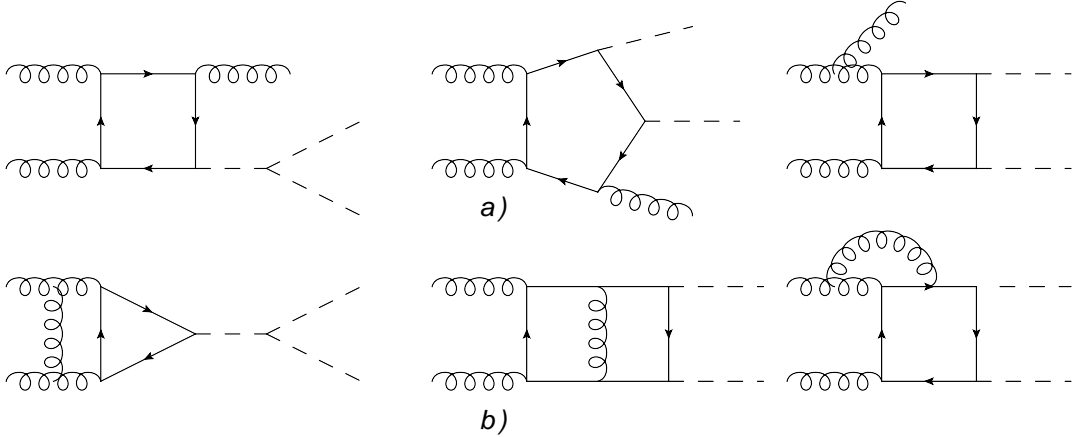}
 \caption{\label{fig:NLO}
Sample of Feynman diagrams for the NLO Higgs pair production in gluon-gluon fusion. a) Real one-loop and b) virtual two-loop corrections.  }
\end{figure*}

In this section we briefly summarise the state-of-the-art in the computation of multiple Higgs production in gluon-gluon fusion.  
For the sake of conciseness we will focus only on Higgs pair production, most of the features being easily extendable to the general case.  
In the SM, the diagrams contributing to the gluon-gluon fusion channel can be organised in two classes:
those where both Higgs bosons couple to the heavy quarks in the loop and those
that feature the Higgs self coupling. The corresponding classes of diagrams appearing at the leading order are shown in fig.~\ref{fig:LO}. 
At NLO, $2 \to 3$ real one-loop diagrams, fig.~\ref{fig:NLO} a), and $2 \to 2$ virtual two-loop diagrams, both boxes and triangles, fig.~\ref{fig:NLO} b), contribute, each of them being infrared divergent yet overall giving a finite result when combined in the computation of collinear and infrared safe observables. The one-loop matrix elements entering the Born amplitude as well as the real corrections can be obtained using modern loop techniques~\cite{Ossola:2007ax} automatically~\cite{Hirschi:2011pa}. The two-loop triangles (such as the first one in fig.~\ref{fig:NLO} b)~) featuring the Higgs self coupling are the same as those entering the single Higgs production and therefore also known for a long time~\cite{Graudenz:1992pv,Spira:1995rr,Harlander:2005rq} and used in publicly available codes for single Higgs production~\cite{Anastasiou:2011pi,Harlander:2012pb}. The two-loop box amplitudes, however, are not known. This computation is extremely challenging due to the presence of several scales ($s,t,u,m_H^2, m_t^2$) in the loops and is currently out of reach.  As it will become clear in the coming section, our reweighting technique can efficiently provide the NLO result with the full top-quark mass dependence once the corresponding amplitudes become known. 
 
In the meantime, precisely due to the difficulty of including higher-order corrections exactly, the strategy widely employed for single Higgs-boson production has been adopted for Higgs pair production. An effective field theory, where the top quark has been integrated out from the theory and the Higgs boson couples directly 
 to the gluon field, has been introduced, where the corresponding lagrangian reads
\begin{equation}
{\cal L}_{\rm HEFT} = \frac{\alpha_S}{12\pi} G^a_{\mu\nu} G^{a, \mu\nu} \, \textrm {log} \left(1+\frac{H}{v}\right),
\label{eq:lag}
\end{equation}
$G$ being the QCD field tensor. The main motivation for using this approximation is that it makes the computation of higher-order corrections feasible.  The approximation has been proven to work extremely well for single Higgs production \cite{Harlander:2009mq}. The HEFT provides accurate predictions for the total rates as well as for the differential distributions when the invariants involved are not much larger than the top quark mass. Unfortunately, in the case of double Higgs production, the relevant scale is at least the invariant mass of the $HH$ pair which is typically $\gtrsim 2 m_t$ and therefore the HEFT provides only a rough approximation for the total rates and a very poor one for the relevant distributions~\cite{Dawson:2012mk,Dolan:2012rv}.

Given the fact that the full NLO results are not presently available and that the HEFT gives a poor description of the process, efforts have been made to improve results taking into account heavy-quark loop effects at least in an approximated way. A first step in this direction has been taken in the seminal NLO calculation for Higgs pair production, as implemented in the code {\sc HPAIR}~\cite{Plehn:1996wb,Dawson:1998py}, which provides total cross sections in the SM and in SUSY. In this case, the NLO calculation is performed within the HEFT, yet all contributions (virtual and real) to the short-distance parton-parton cross section are expressed in terms of the LO  cross section times an $\alpha_S$ correction. 
The LO cross section in the HEFT is then substituted by the LO one with the full heavy-quark mass dependence. 
As the aim of {\sc HPAIR} is to provide results for the total cross section, this approximation is certainly a first useful step. However,  such an approach is obviously not suitable
 in general and in particular for observables that receive considerable contributions from the real $2 \to 3$ configurations, the most glaring example being the tail of the transverse momentum of the Higgs pair, $p_T(HH)$.

We also mention that the top-quark mass effects at NLO in QCD have been estimated through an $1/m_t^2$ expansion \cite{Grigo:2013rya} (on which we will further comment later). The NNLO computation for total rates in the HEFT~\cite{deFlorian:2013jea} has also now been completed with the calculation of the three-loop matching coefficient $C_{HH}$ \cite{Grigo:2014jma}, while recently results have been obtained by merging samples of different parton multiplicities in~\cite{Li:2013flc,Maierhofer:2013sha} and including threshold resummation \cite{Shao:2013bz}.

\section{The inclusion of heavy-quark loop effects }
\label{sec:method}

In this section we discuss how to improve further on the HEFT approximation,
and in particular how one can improve on pure NLO in HEFT by including
heavy-quark loop effects in the Born amplitude as well as in the real emission ones. Results using this method have already
 been presented in \cite{Frederix:2014hta}. They have been achieved by implementing all the ingredients
in an automatically generated code for the HEFT within the {\sc MadGraph5\_aMC@NLO}
framework~\cite{Alwall:2014hca}. In practice, the effective Lagrangian
of eq.~(\ref{eq:lag}) is implemented in {\sc FeynRules} \cite{Christensen:2008py,Alloul:2013bka}\footnote{More complete implementations including interactions coming from the full set of dimension-six operators are also available~\cite{Artoisenet:2013puc,Alloul:2013naa,Maltoni:2013sma,Demartin:2014fia}.}  upgraded to NLO by including the necessary UV and $R_2$ counter terms~\cite{Ossola:2006us,Ossola:2007ax, Degrande:2014vpa,Page:2013xla,Demartin:2014fia} to obtain a UFO model \cite{Degrande:2011ua,deAquino:2011ub} that can be imported in {\sc MadGraph5\_aMC@NLO}.  At this point NLO parton--level events can be automatically generated in the HEFT and then reweighted on an event--by--event basis using the appropriate  heavy-quark loop amplitudes.

To include the heavy-quark amplitudes we make use of the structure of an NLO calculation as performed within {\sc MadGraph5\_aMC@NLO}. For completeness, 
let us consider the computation of a cross section at NLO, namely
 \begin{equation}
     d\sigma = d\phi_n \left( \mathcal B + \mathcal V + \mathcal C^{int} \right) +
     d\phi_{n+1} \left(\mathcal R - \mathcal C\right)\,,
     \label{eq:nloxsec}
 \end{equation}
 where $\mathcal{B,V,R}$ are respectively the Born, virtual and real emission contributions, $\mathcal C$ are the local
 counterterms (needed in order to render the integral over $d\phi_{n+1}$ finite) and $\mathcal C^{int}$ is the integrated form of  $\mathcal C$ (over the 
 extra parton phase space). The detailed form of the counterterms $\mathcal C^{(int)}$ depends on the subtraction scheme in use for 
 the computation ({\it e.g.},  FKS\cite{Frixione:1995ms} or CS\cite{Catani:1996vz}).
In general they involve a Born matrix element times some 
 process-independent splitting kernel together with a dedicated phase-space mapping.   
 
  
     A very similar formulation as the one above holds in the case of matching NLO computations with parton-shower
     in the MC@NLO formalism, where on top of the local counterterms $\mathcal C$ one has to also include the so-called Monte-Carlo counterterms  $\mathcal C_{MC}$~\cite{Frixione:2002ik}.  The important difference with respect to the NLO formulation of eq.~(\ref{eq:nloxsec}) is that the MC counterterms are such that Born-like ($\clS$-events) and real-emission ($\clH$-events) unweighted events can obtained as the corresponding subtracted cross sections are separately finite. The corresponding contributions to the total cross section can be written as
 \begin{eqnarray}
     d\sigma ^{(\clH)} & = & d\phi_{n+1} \left( \mathcal R - \mathcal C_{MC} \right)\,, \\ 
     d\sigma ^{(\clS)} & = & d\phi_{n+1} \left[ \left( \mathcal B + \mathcal V + \mathcal C^{int} \right) \frac{d\phi_{n}}{d\phi_{n+1}}
     + \left( \mathcal C_{MC} - \mathcal C \right) \right]\,. 
 \end{eqnarray}
     In the {\sc MadGraph5\_aMC@NLO} framework, one can automatically generate the code corresponding to the Born, virtual, real amplitudes, the counter terms and the phase space~\cite{Frederix:2009yq,Hirschi:2011pa}  in one go in order to compute cross sections and generate events for  $gg\to HH$ at NLO in QCD in the HEFT. All the finite heavy-quark one-loop matrix-elements ({\it i.e.} those entering the Born and real contributions) needed can also be obtained within {\sc MadGraph5\_aMC@NLO}.
 Note, however, that two limitations presently make the {\it automatic} computation of the exact NLO result not possible. First, the computation of cross sections that have a loop Born matrix-element is not automated yet (even at the LO only). Second, even with the automation for loop-induced processes, the need for the two-loop amplitudes would require an external routine, as this cannot be performed automatically by {\sc{MadLoop}}.
Therefore, the inclusion of heavy-quark effects needs manipulation that can in principle be performed in two ways. 
 
The first option is to generate the code for an NLO computation in the HEFT and then replace the matrix-elements 
             (for $\mathcal {B,V,R},\mathcal C^{int}$ and $\mathcal C_{MC}$) with the corresponding  ones in the FT. 
             Even though this is the simplest option, it features several drawbacks. First, this method is very
             inefficient as the (computationally expensive) one-loop and two-loop matrix elements routines would then be called many times to probe and map all regions
             of phase space. In addition, it requires the evaluation of the real one-loop matrix elements in the FT in regions of phase space very close to the soft/collinear limits, {\it i.e.} where they might feature unstable configurations. For such points, multiple precision needs to be employed at the cost of a growth of the running time by a factor of a hundred.
 
 The second option is to include the top-quark mass effects by reweighting after having generated the short-distance events  and before these are passed to a parton shower program. In order for this procedure to be applied, all the weights corresponding to the
         separate contributions (events and counter events)  and the corresponding kinematics, which is in general different between events and each of the counter events, need to be  saved in an intermediate event file. With this information it is then possible to recompute the total event weight by reweighting each
         contribution by the matrix-elements in the FT. The weights corresponding to $\mathcal{B,V,C}^{(int)},\mathcal C_{MC}$ are rescaled by the ratio  $\mathcal B_{FT} / \mathcal B_{HEFT}$, while those corresponding to $\mathcal R$ by the ratio  $\mathcal R_{FT} / \mathcal R_{HEFT}$. When
         unweighted events are generated, this amounts into rescaling the whole weight of $\clS$-events with Born matrix-elements, and
         the different terms corresponding to $\clH$-events as written above. This solution has the advantage of requiring the FT matrix-elements to be evaluated in significantly fewer phase space points than those used while integrating it directly. In addition, it is completely general and only assumes that there are no regions in phase space where the HEFT gives a vanishing contribution while the full theory does not. In our case this condition is satisfied. Moreover, being non-renormalisable, the HEFT leads to distributions that are in general harder in the high-energy tails than the FT. Therefore using the HEFT to generate events efficiently populates phase space regions that are later suppressed by the FT matrix elements, implying no large degradation of statistical accuracy.
              
 In this work we follow the latter strategy, which we find efficient and general.  Such technique 
 has been proposed and employed to include heavy-quark corrections at LO in multi-parton merged samples for single Higgs \cite{Alwall:2011cy} 
 and $HH$ \cite{Li:2013flc} production. 
 The implementation of the reweighting is simplified considerably by the fact that {\sc MadGraph5\_aMC@NLO} 
already features the needed structure and automatically stores all the relevant weights 
(though in a slightly different form than the one presented in eq.~(\ref{eq:nloxsec})), 
in order to evaluate the uncertainties associated with scale variation and PDF on an event-by-event basis as described in~\cite{Frederix:2011ss}.

If the two-loop matrix elements were known, our procedure would have allowed to exactly compute cross sections at NLO accuracy and
perform event generation at NLO+PS in the FT. As the finite part of the two-loop box virtual correction is unknown, this is not yet possible.
In our work, we therefore employ and study the effect of using different approximations for the only unknown terms, while including all 
one-loop (Born and real) known terms. Once the two-loop results for the virtual corrections become available,
it will be possible to include them straightforwardly, for example following the conventions of the 
Binoth Les Houches Accord \cite{Alioli:2013nda}. 
 
The method described here can be applied to other loop-induced processes, for which suitably defined effective interactions
can be used to automatically generate events while the FT one--loop matrix elements can then be employed to correctly 
reweight the events. As already mentioned, our method has been already used  in the NLO calculation of Higgs pair production within the SM in \cite{Frederix:2014hta}
 and more recently in the 2HDM in \cite{Hespel:2014sla}. In this work, we will also apply the same procedure to triple Higgs production. We stress that it could also be applied to other loop-induced processes, such $H+1$ jet and $HZ$ associated production in gluon-gluon fusion.

\section{Higgs pair production}

In this section we present the results obtained by applying the method described in the previous section to $HH$ production. 
The first effects we wish to discuss are those coming from the inclusion of the top-quark width.
The relevance of the top-quark width in the context of single Higgs production in gluon-gluon fusion has been discussed in \cite{Anastasiou:2011pi} 
 where it has been verified that for a light Higgs below the $t\bar{t}$ threshold the effect is negligible
 but rises to the percent level for heavier Higgs bosons, {\it i.e.}, closer to the $t \bar t $ threshold.  Such virtualities are exactly 
 those relevant for Higgs pair production and one can therefore expect the top-quark width effects
not to be negligible. 

\begin{figure}[t!]
 \center 
 \includegraphics[scale=0.8]{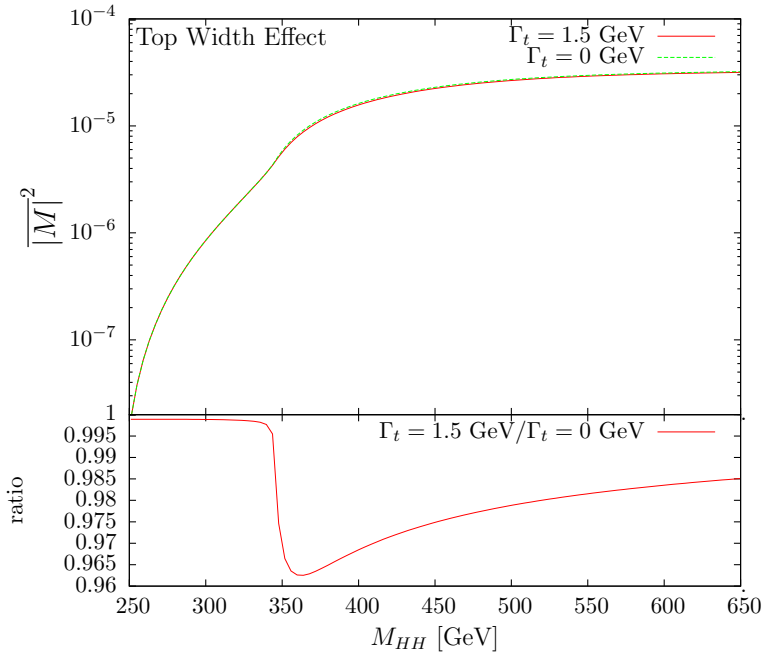}
 \caption{\label{fig:CMS}
Top width effect on the one-loop (Born) matrix element squared for $g g \rightarrow H H$. 
The results for $\Gamma_t=0$ and 1.5 GeV are shown along with the corresponding ratio.}
\end{figure}

As already mentioned, in our calculation the $2\to 2$ and $2 \to 3$ one-loop amplitudes are obtained via {\sc MadLoop}~\cite{Hirschi:2011pa}. The computation is performed using the complex mass scheme for the top quark~\cite{Denner:1999gp,Denner:2005fg} as  implemented in {\sc MadLoop} and {\sc ALOHA}  in {\sc MadGraph5\_aMC@NLO}. In practice, for the top quark this simply amounts to performing the replacement 
\begin{equation}
m_t \to m_t \sqrt{1 - i \Gamma_t/m_t}
\end{equation}
everywhere in the computation, {\it i.e.} to modify the kinematical mass as well as the Yukawa coupling. 
The effect of including a non-zero top-quark width is shown in  fig.~\ref{fig:CMS}, where the LO matrix
element squared for $g g \rightarrow H H  $ is plotted as a function of the invariant mass of the Higgs pair for $\Gamma_t=0$ and 1.5 GeV.\footnote{We note that here we assumed a 90$^{\circ}$ scattering for all points included in fig.~\ref{fig:CMS}, 
but as the matrix element has an extremely weak angular dependence~\cite{Dawson:2012mk}, this provides a perfectly good 
demonstration of the effect also at the level of the Higgs pair invariant mass distribution.}  A behaviour similar in size and with the same negative sign as the single Higgs case~\cite{Anastasiou:2011pi} is found, with the non--zero width result displaying a maximum decrease of $\sim$4\% compared to the narrow width result right after the $t\bar{t}$ threshold. 
 The results shown here have been obtained at the matrix element squared level. 
The final effect on the total cross section at LO at 14 TeV LHC is shown in tab.~\ref{tab:14TeV} and amounts to
a correction of $\sim$-2\%. For our NLO predictions we will use a top-quark width of 1.5 GeV.

We now consider the inclusion of the finite top mass in the NLO computation. In what we dub NLO FT$_{{\rm approx}}$, the Born and real configurations are reweighted with the corresponding Born and real emission finite top-quark mass matrix elements and for the virtual configurations, the HEFT result, yet rescaled by the Born in the FT, is used. We stress again that the only approximation made in this procedure is that coming from the absence of the exact results for the two-loop virtual terms. As a check we have applied this method to single Higgs production in gluon-gluon fusion where all elements of the calculation are available and found excellent agreement with the corresponding FT implementation  in {\sc MC@NLO v4.0}~\cite{Frixione:2010wd}. 

We also compare our NLO FT$_{{\rm approx}}$ results to two other approximations. 
The first of these two approximations corresponds to what in the following we refer to as ``Born-improved'' HEFT results and amounts to rescaling all events, Born and real-emission like, by the Born in the FT, {\it i.e.}, to always use the weight $\mathcal B_{FT} / \mathcal B_{HEFT}$. In the case of the real-emission events the kinematics used is that determined by the counter terms mapping. This approach mimics the one used in {\sc HPAIR}, the only (practically irrelevant) difference being that in {\sc HPAIR} the reweighting is performed after the integration over the polar angle, while our HEFT Born reweighting is fully differential. Indeed a detailed check with the results of {\sc HPAIR} shows that once the same parameters are chosen, and in particular the top-width is set to zero, an excellent agreement is found.

In addition to the Born-improved HEFT results, we compare our reference results to a second approximation, 
dubbed NLO FT$'_{{\rm approx}}$, by which we assess 
the possible shifts of the central value of the NLO cross section due to the 
inclusion of the exact two-loop corrections in the FT. As already mentioned several times, the two-loop box contributions, {\it e.g.}, those coming from the second and third diagrams of  fig.~\ref{fig:NLO}b)  are not known. However, the two-loop triangle contributions, {\it e.g.}, first diagram of fig.~\ref{fig:NLO}b),  are known (and form a gauge independent subset) as they are the same as those entering single Higgs production. To this aim, we use the finite contribution of the two--loop corrections for single Higgs production, which are publicly available as a function of the Higgs mass, {\it e.g.} in {\sc SusHi} \cite{Harlander:2012pb}, 
and assume that the same form factor would hold for the two-loop box amplitude. In other words, we proceed under the assumption that these corrections factorise out of the matrix element globally, {\it i.e.} for both the box and the triangle diagrams. Needless to say, such an assumption is \emph{ad hoc} and while we do not claim that the triangle form-factor should resemble the box one, this test is still a useful one. The main reason comes from the fact that it allows us to study the extent of possible cancellations taking place between the top-quark mass corrections from the real (which are included) and virtual (which are only approximate) matrix elements. It can be observed\footnote{We are grateful to Michael Spira for bringing up this point.} that in single Higgs production, where the exact NLO results are available,  the ``Born-rescaled'' HEFT result is extremely close to the exact value for a Higgs mass near the $t\bar{t}$ threshold \cite{Harlander:2003xy,Anastasiou:2009kn}. It is therefore important to assess whether a similar effect might be taking place also in $HH$ production. Such a cancellation would be spoiled in our NLO FT$_{{\rm approx}}$, where only the exact real emission effects and not those coming from the two-loop virtual contributions are included. As we will show in the following, this does not seem to be the case. 

We quantify the differences between the approximations discussed above by calculating the total cross section for Higgs pair production
in gluon-gluon fusion at 14 TeV in tab. \ref{tab:14TeV}. In this computation, we have set the Higgs mass to
$m_H=125$~GeV and the top-quark mass to $m_t=172.5$~GeV.  We note that the top-quark mass dependence of the cross section around the reference top mass of 172.5 GeV can be parametrised approximately (at LO FT as well as at NLO FT$_{\rm approx}$) by  $\Delta\sigma/\sigma\sim0.6\% \times \Delta m_t/$GeV. Parton distribution functions (PDFs) are evaluated by
using the MSTW2008 (LO and NLO) parametrisation in the five-flavour scheme~\cite{Martin:2009iq}. 
The renormalisation and factorisation scales $\mu_{R,F} $ are set to $\mu_{R}=\mu_{F}=\mu_0=m_{HH}/2$. 
The dependence of the predictions on scale and PDF variations can be estimated at no extra computational cost via a reweighting technique~\cite{Frederix:2011ss}. 
Scales are varied independently in the range $\mu_0/2 < \mu_R, \mu_F < 2\mu_0$ and PDF uncertainties at the 68\% C.L. are obtained following the prescription 
given by the MSTW collaboration~\cite{Martin:2009iq}. Even though $b$-quark loops can be computed in our setup, $b$-quark masses as well as their tiny ($\sim$0.3\%) contribution to the $HH$ cross section are neglected in the following. 

Table~\ref{tab:14TeV} collects our results. We first verify that the effect of the non--zero top-quark width on the total cross section at LO, a $\sim 2$\% decrease, directly follows from the results shown in fig.~\ref{fig:CMS} and the fact that the invariant mass distribution peaks at $\sim$400~GeV. We also note the well-known fact that the process receives large QCD corrections as well as the expected reduction of the theoretical uncertainties for the NLO computations.  We then show three NLO results: i) the Born-improved HEFT result through a local event-by-event reweighting, ii) the NLO FT$_{\rm {approx}}$ result, obtained by combining the exact real emission matrix elements, with the Born-rescaled HEFT results for the virtual corrections and iii)
the NLO FT$'_{\rm approx}$ result obtained by combining the exact real emission matrix elements, with the exact results of single Higgs production for
the virtual corrections, as described previously. For all NLO results we keep the finite top-quark width of 1.5 GeV.

\begin{table}[]
\renewcommand{\arraystretch}{1.3}
\begin{center}
 \begin{tabular}{c|l|c} \hline \hline
 \multicolumn{2}{l|} {$HH$ production in gluon-gluon fusion at 14 TeV} & Cross section [fb] \\ \hline
&HEFT  &  19.2$^{+35.2+2.8\%}_{-24.3-2.9\%}$ \\ \cline{2-3}
LO & FT, $\Gamma_t=0$ GeV & 23.2$^{+32.3+2.0\%}_{-22.9-2.3\%}$  \\  \cline{2-3}
& FT, $\Gamma_t=1.5$ GeV  & 22.7$^{+32.3+2.0\%}_{-22.9-2.3\%}$  \\  \hline
\multirow{4}{*}{NLO} & HEFT  &  32.9$^{+18.1+2.9\%}_{-15.5-3.7\%}$\\  \cline{2-3}
& HEFT Born-improved &  38.5$^{+18.4+2.0\%}_{-15.1-2.4\%}$ \\  \cline{2-3}
 & FT$_{\rm approx}$  (virtuals: Born-rescaled HEFT ) & 34.3$^{+15.0+1.5\%}_{-13.4-2.4\%}$  \\ \cline{2-3}
 & FT$'_{\rm approx}$  (virtuals: estimated from single Higgs in FT) & 35.0$^{+15.7+2.0\%}_{-13.7-2.4\%}$ \\ \hline \hline
 \end{tabular}
 \end{center}
 \caption{Cross section results (in fb) for Higgs pair production in gluon-gluon fusion at 14~TeV. LO results in the Full Theory are given without and with top-quark width effects. The first NLO result corresponds to the HEFT, while the second to the Born-improved HEFT. The third NLO result, FT$_{\rm approx}$, corresponds to our baseline approach where all known top-quark mass corrections coming from one-loop amplitudes are included and the HEFT Born-rescaled approximation for the two-loop amplitudes is used. In the last result, FT$'_{\rm approx}$ , the information from the known two-loop triangles is also used to estimate the full two-loop contributions. More details are given in the text. All NLO results feature a finite top-quark width. The first uncertainty quoted refers to scale variations, while the second to PDFs. Uncertainties are in percent. No cuts are applied to final state particles and no branching ratios are included. }
 \label{tab:14TeV}
\end{table}

We can now compare the different approximations of the FT NLO result. The first important observation is that
the Born-improved result is 11\% larger than our baseline one. We also note here that the 
Born-improved result obtained by a local event-by-event rescaling is within 1\% of what one would obtain from a global Born rescaling obtained using the total cross section numbers, {\it  i.e.}, $\sigma^{NLO}_{HEFT}\times \sigma^{LO}_{FT}/\sigma^{LO}_{HEFT}$. The difference from the Born-improved result only slightly reduces (9\%) when an estimate for
the finite top-quark mass terms from the two-loop contributions is included, see last line of tab.~\ref{tab:14TeV}. Our NLO FT$_{\rm approx}$ result is rather stable in that respect. 
This is related to the fact that the cancellation we discussed earlier for single Higgs production is only relevant very close to the $t\bar{t}$ threshold, with the Born-rescaled result
 rapidly rising over the exact one above the threshold. 
In the case of single Higgs production, we have indeed checked that for Higgs masses above 400 GeV the NLO FT$_{\rm approx}$ result (only including the exact real emission matrix element but not the known two--loop virtual results) is closer to the exact result than the corresponding Born-improved one. In the case of Higgs pair production, one could also argue that even if a similar cancellation of the top-quark mass effects between the real and virtual corrections occurred at the $t\bar{t}$ threshold, it would not have a very pronounced effect on the total cross section, as for Higgs pair production the peak of the invariant mass distribution is located at higher mass values. 

At this point it is worth to recall the results of ref.~\cite{Grigo:2013rya}, where the top-quark mass effects at NLO in QCD
were estimated by computing the first few terms in the $1/m_t^2$ expansion for the $K-$factor. The $1/m_t^2$ expansion is known not to converge well at LO~\cite{Dawson:2012mk} and is not supposed to work beyond or even close to the $\sqrt{s}=2 m_t$ threshold, around and beyond which the bulk of the $HH$ cross section resides. However, in ref.~\cite{Grigo:2013rya} an attempt was made by combining the exact Born cross section with the $1/m_t^2$ expanded $K-$factors, as a ``taming"  technique for the expansion. A +10\% 
 increase with respect to the Born-rescaled HEFT result was found, {\it  i.e.}, an effect similar in size but opposite in sign to our estimate. Combined with our findings, the estimate of ref.~\cite{Grigo:2013rya} implies that the difference between the finite part of the Born-rescaled HEFT virtuals and the exact ones should account for a +20\% increase of the total cross section, a quite large effect indeed, especially considering that by including top-mass effects in the virtual corrections estimated via the known two-loop triangles, leads only to a couple of percent increase. Besides, we note that the results of the same $1/m_t^2$ expansion approach applied to the production of a single heavy Higgs of mass between 400 and 500 GeV, are known to overestimate the exact results in the FT when no high-energy matching is performed \cite{Harlander:2002wh,Harlander:2012pb,Harlander:2009my}. 
 
While only an exact calculation of the missing two-loop amplitudes will finally settle this issue, the NLO FT$_{\rm approx}$ approach provides central values for the cross sections that appear rather robust, predicting a correction of about -10\% with respect to those obtained by means of the Born-improved HEFT. In addition, together with the results of ref.~\cite{Grigo:2013rya}, our study provides an estimate of about 10\% for the uncertainty to be associated with the HEFT calculation due to the missing top-quark mass effects. Such an uncertainty should be quoted along with the other theoretical uncertainties in the HEFT calculations, at NLO but also at NNLO. 

\begin{figure}[!th]
 \center 
\hspace*{.1cm} \includegraphics[scale=0.66]{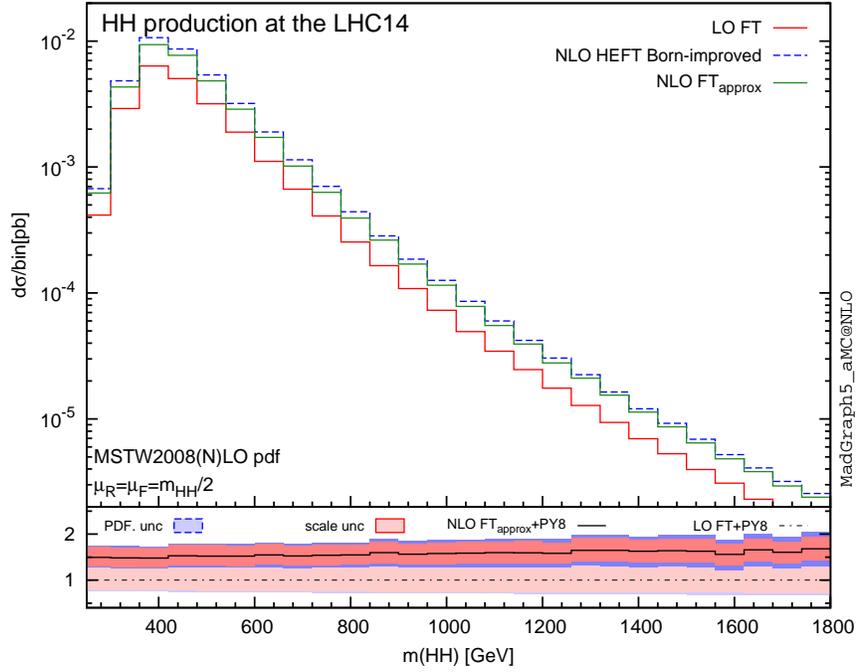}
 \vskip1cm
\hspace*{.4cm} \includegraphics[scale=0.66]{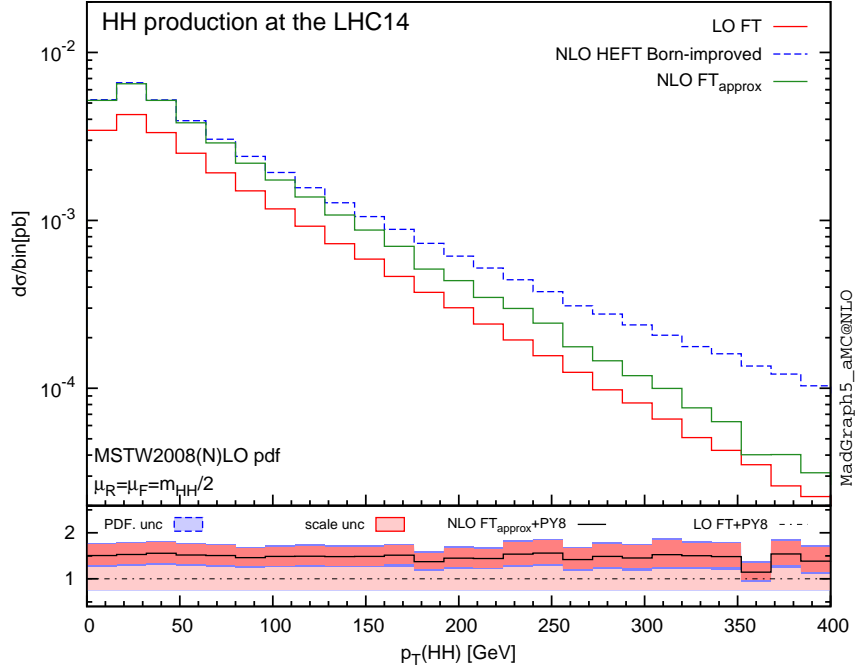}
 \caption{\label{fig:mhhpthh}
Differential cross sections for the invariant mass and transverse momentum of the Higgs pair in $HH$ production via gluon-gluon fusion at the LHC with 
 14 TeV centre-of-mass energy. Distributions are obtained by generating events at parton level at LO and NLO accuracy and then matching 
to {\sc Pythia8}. Uncertainties corresponding to PDF and scale variations are shown in the lower inset.  }
\end{figure}

Finally, we note that including the exact one--loop $2 \to 3$ matrix elements provides a more accurate description of the tails
of the distributions where hard parton emissions take place, and the factorisation of the $2 \to 3$ real--emission amplitudes into 
the $2 \to 2$ LO amplitudes, as implicit in the Born-improved approach, cannot accurately describe the hard parton kinematics.
To emphasize this point,  we compare  in fig.~\ref{fig:mhhpthh} the differential distributions of the invariant mass  
and the transverse momentum of the Higgs pair obtained by NLO FT$_{\rm approx}$ and Born-improved HEFT. 
{\sc Pythia8}~\cite{Sjostrand:2007gs} has been used for parton shower and hadronisation.

By studying the two distributions, we see that including the exact real emission matrix elements affects the two observables in different ways. 
On the one hand, for the invariant mass distribution the effect is a uniform modification of the cross section by about 10\%. On the other hand, 
for the transverse momentum of the Higgs pair, at low values where the distribution is dominated by the parton shower, there is no visible difference
between the two results, while at high $p_T$ values where the distribution is dominated by hard parton emission, coming from the real matrix elements,
we see that there is a significant deviation from the Born-improved result. In that region we trust our FT$_{\rm approx}$ prediction more as
it describes better the kinematics.

\begin{figure}[t!]
 \center 
\hspace*{.4cm} \includegraphics[scale=0.9]{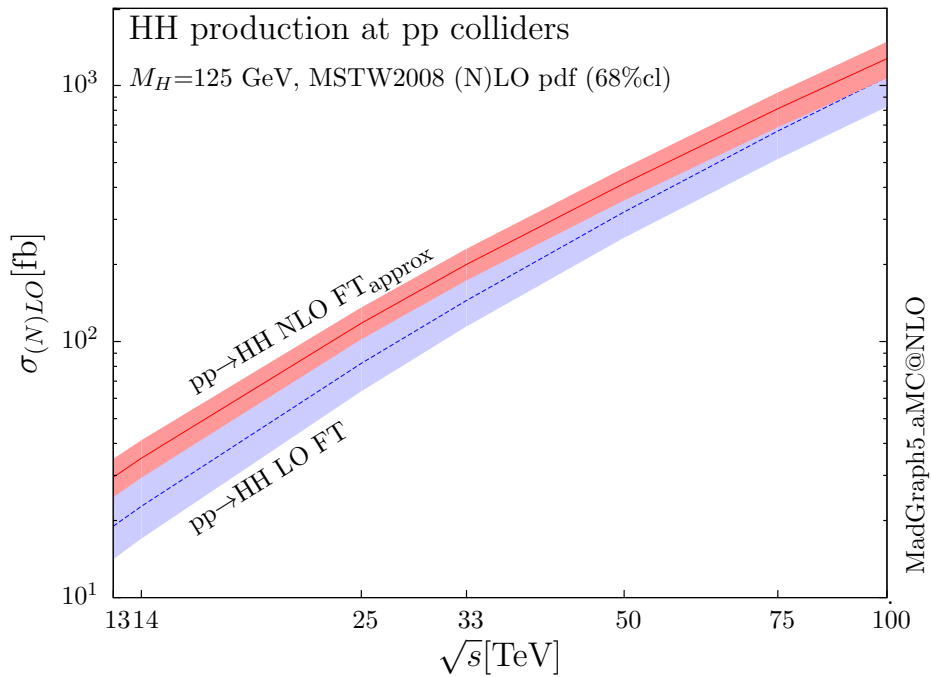}
 \caption{\label{fig:ggxsec}
Total cross sections at LO and NLO in QCD in the FT for the Higgs pair production from gluon-gluon fusion at $pp$ colliders as a function of the centre-of-mass energy. The thickness of  the lines corresponds to the scale variation and PDF uncertainties added linearly. }
\end{figure}

\begin{table*}[t]
\renewcommand{\arraystretch}{1.3}
\begin{center}
    \begin{tabular}{l|c|c|c}
        \hline \hline
     $\sigma(HH)$ [fb]    & $\sqrt{s} = 14$~TeV
         & $\sqrt{s} = 33$~TeV
         & $\sqrt{s} = 100$~TeV\\
         \hline
         LO FT&  
 22.7 $^{+32.3+2.0\%}_{-22.9-2.3\%}$ & 
145 $^{+24.9+1.2\%}_{+18.9-1.9\%}$ & 1080 $^{+28.6+1.0\%}_{-21.8-1.7\%}$ \\         
        NLO FT$_{\rm approx}$ &  34.3$^{+15.0+1.5\%}_{-13.4-2.4\%}$ & 199 $^{+13.2+1.3\%}_{-11.6-1.6\%}$
& 1250 $^{+14.8+1.0\%}_{-14.4-1.6\%}$ \\
\hline \hline
\end{tabular}

    \caption{\label{tab:HHsec} LO and NLO total cross sections (in fb) for Higgs pair production
    in gluon-gluon fusion at $\sqrt{s} =  14, 33, 100$~TeV in the FT. 
      The  uncertainties (in percent) refer to scale variations and to PDF, respectively.  
      No cuts are applied to final state particles and no  branching ratios are included.}  
\end{center} 
\end{table*}

For completeness, we also show the results for the total cross section at LO and NLO for Higgs pair production in gluon-gluon fusion
as a function of the hadronic centre-of-mass energy in fig.~\ref{fig:ggxsec}, including the  uncertainties due to scale variation and PDF added linearly.
The uncertainty bands demonstrate nicely the reduction of the scale uncertainties for the NLO calculation. 
Results for a selected number of hadronic energies are also shown in tab. \ref{tab:HHsec}. 

\section{Higgs triple production}

\begin{figure*}[tb]
 \center 
 \includegraphics[scale=0.6]{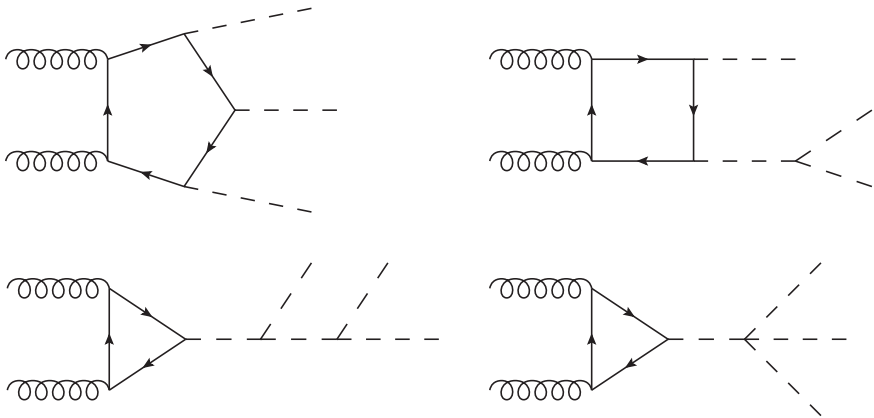}
 \caption{\label{fig:hhh}
Representative Feynman diagrams for triple Higgs production in gluon-gluon fusion. }
\end{figure*}

We now apply the same procedure followed for Higgs pair production to obtain results beyond the LO
for triple Higgs production through gluon-gluon fusion.  Representative topologies of diagrams that contribute to the process at LO
 are shown in fig.~\ref{fig:hhh}. These include pentagons where the Higgs bosons couple only to the heavy quarks of the loop, boxes in combination with one power 
of the trilinear Higgs coupling and triangles with either two powers of the trilinear coupling or the quartic Higgs coupling. It turns out
that at LO, the hierarchy of the various contributions in decreasing size order is first pentagons, then boxes and finally triangles~\cite{Plehn:2005nk}. 
With the triangles contribution already being the smallest of the three, and the fact that the quartic coupling itself has only a very small contribution to that
(it only appears in one of the four diagrams), any attempt  to access the quartic coupling through the measurement of this process
becomes not feasible at the LHC and extremely challenging even at a future 100~TeV collider. In any case, measuring the quartic Higgs coupling in this
process would require a very precise knowledge of the triple Higgs coupling \cite{Plehn:2005nk}.  The situation could possibly change in the presence of physics beyond the SM. In this respect having predictions as accurate as possible for production cross sections at hadron colliders in the SM is still be very useful. 

We therefore propose to follow the same procedure as for Higgs pair production and improve the NLO HEFT results by systematically including
the information from the FT one-loop amplitudes, the Born and the real contributions. In this case, the information which will be included 
in an approximated way is that related to the two-loop boxes and two-loop pentagons, whose calculation is extremely challenging (note that even the two-loop boxes are more complicated than those in $HH$ production as they feature one more scale) and probably will not be available for some time. 
The method and the setup follows exactly that of $HH$, even though the calculation is more complicated and the resulting reweighting procedure is substantially slower. For this process, we find it necessary to use a small computing cluster and fully parallelise the reweighting on an event-by-event basis.

\begin{figure}[t!]
 \center 
 \includegraphics[scale=0.9]{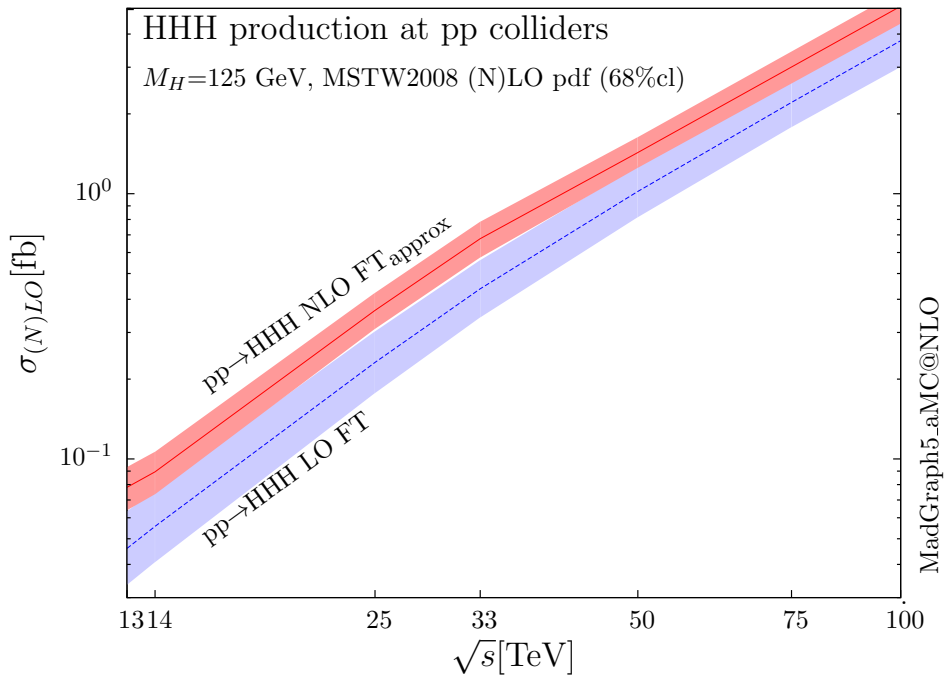}
 \caption{\label{fig:HHHxsec}
Total cross sections at LO and NLO in QCD in the FT for the triple Higgs production from gluon-gluon fusion at $pp$ colliders as a function of the centre-of-mass energy.
The thickness of  the lines corresponds to the scale and PDF uncertainties added linearly. }
\end{figure}

We show our results for the production cross sections as a function of the centre-of-mass energy in fig.~\ref{fig:HHHxsec} 
and a few representative results in tab.~\ref{tab:HHHsec}. The NLO calculation leads to $K-$factors and uncertainties which are similar to Higgs pair production. 
The most important information conveyed by fig.~\ref{fig:HHHxsec} is that the cross section remains very small even at 100~TeV $pp$ collisions. 

We conclude by showing the NLO+PS effects in two key distributions, {\it  i.e.} that of the invariant mass of the three Higgs-bosons and the transverse
momentum of the triplet. The latter distribution features two important damping effects: at small  $p_T(HHH)$ the spectrum is softened by the soft
resummation performed by the shower and at high $p_T(HHH)$ where the top-quark loop effects matter and the HEFT is not reliable. 

\begin{table*}[t]
\renewcommand{\arraystretch}{1.3}
\begin{center}
    \begin{tabular}{l|c|c|c}
        \hline \hline
     $\sigma(HHH)$ [fb]     & $\sqrt{s} = 14$~TeV
         & $\sqrt{s} = 33$~TeV
         & $\sqrt{s} = 100$~TeV\\
         \hline
         LO FT&  
 0.0557 $^{+34.5+2.5\%}_{-24.0-2.7\%}$ & 
0.438 $^{+26.8+1.5\%}_{+20.0-2.0\%}$ & 3.78 $^{+24.1+0.9\%}_{-18.7-1.7\%}$ \\         
        NLO FT$_{\rm approx}$ &  0.0894$^{+16.5+2.5\%}_{-14.6-3.2\%}$ & 0.677 $^{+14.5+1.4\%}_{-13.4-1.7\%}$
& 5.09 $^{+13.5+1.0\%}_{-12.7-1.3\%}$ \\
\hline \hline
\end{tabular}

    \caption{\label{tab:HHHsec} LO and NLO total cross sections (in fb) for triple Higgs production
    in gluon-gluon fusion at $\sqrt{s} =  14, 33, 100$~TeV in the FT. 
     The  uncertainties (in percent) refer to scale variations and to PDF, respectively.  No cuts are applied to final state particles and no
        branching ratios are included.}  
\end{center} 
\end{table*}

 \begin{figure}[h!]
  \center 
\hspace*{.5cm}  \includegraphics[scale=0.66]{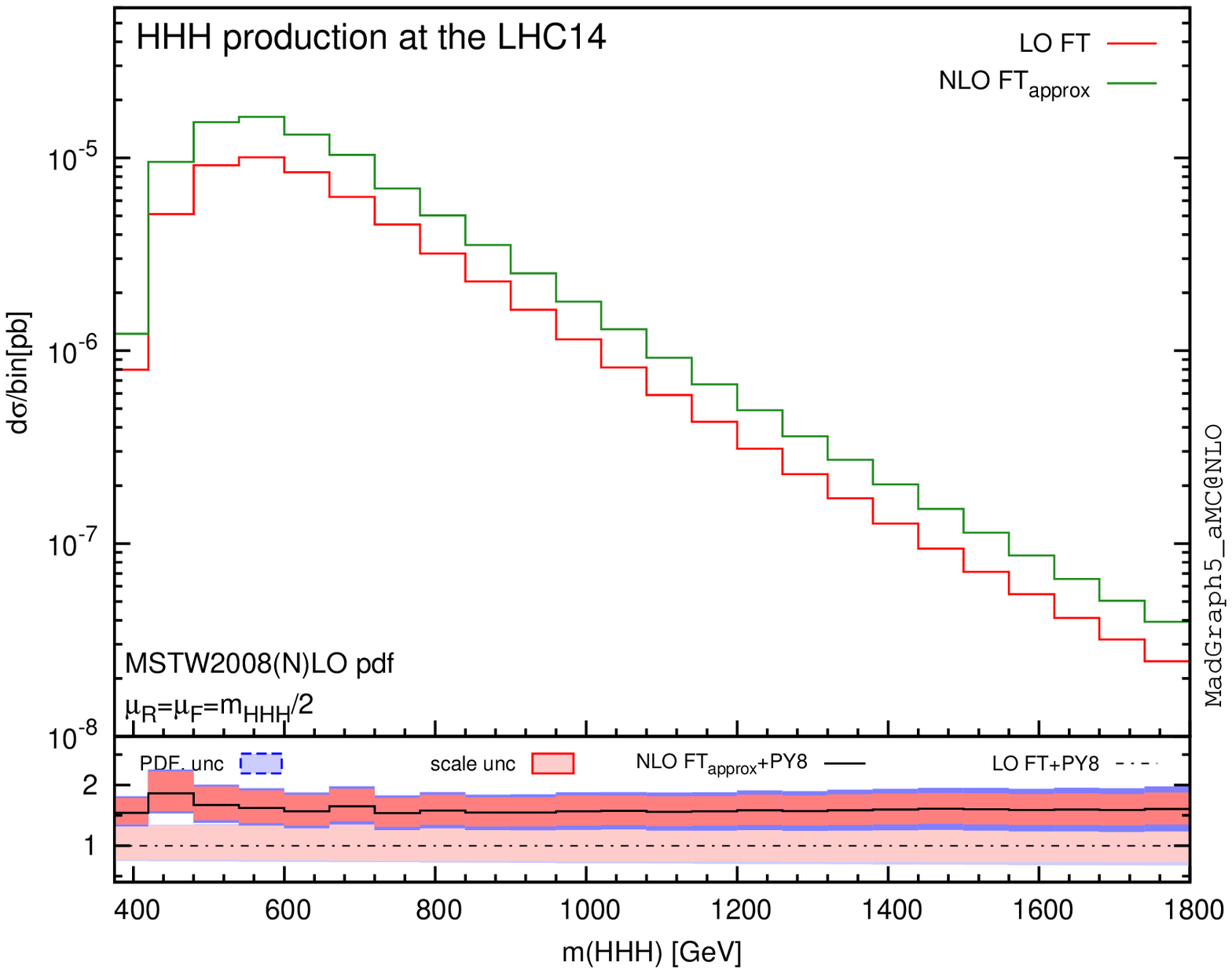}
  \vskip1cm
\hspace*{.5cm}  \includegraphics[scale=0.66]{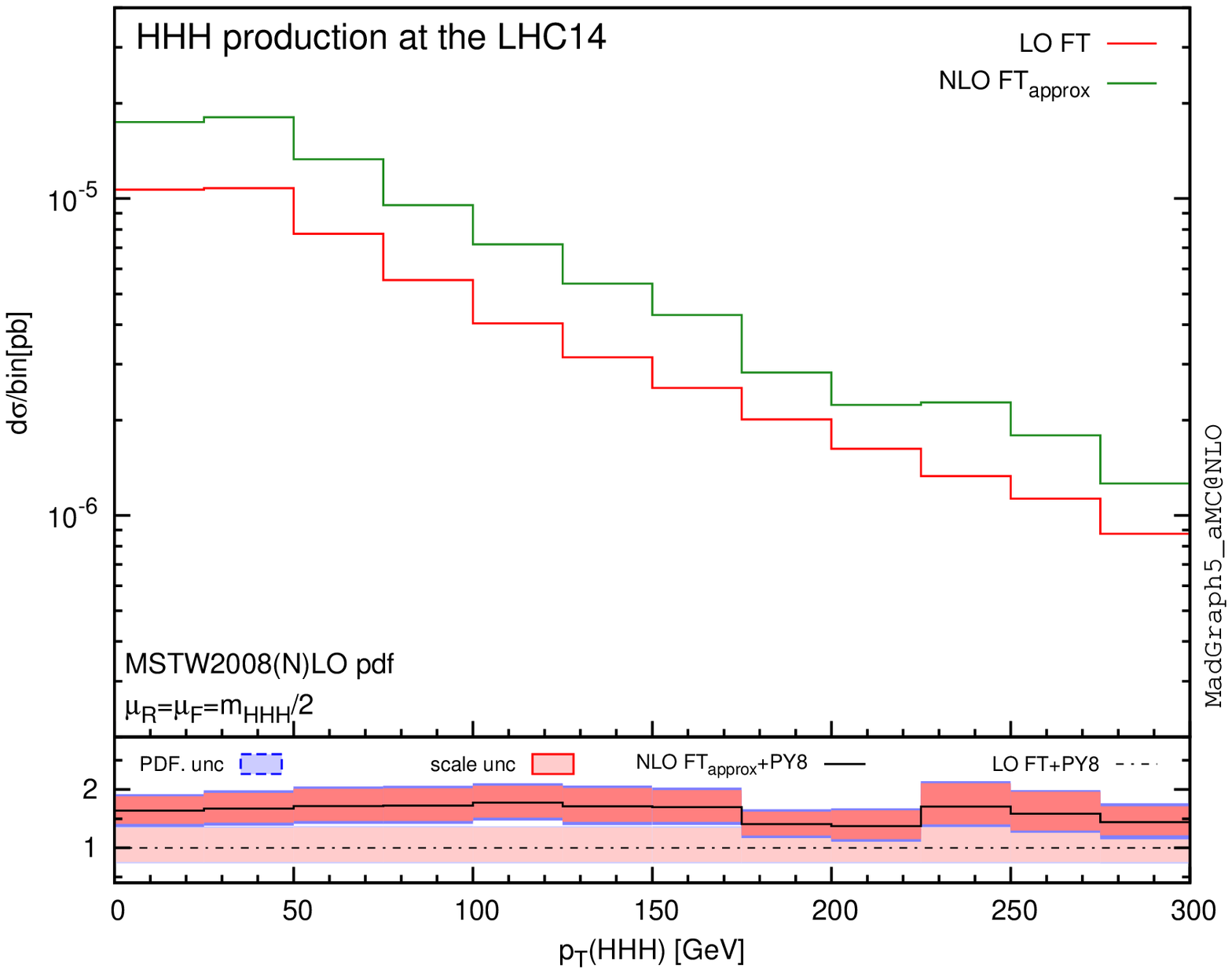}
  \caption{\label{fig:mhhhpthhh}
 Differential cross sections for the invariant mass and transverse momentum of the three Higgs system in $HHH$ production via gluon-gluon fusion at the LHC with 
  14 TeV centre-of-mass energy. Distributions are obtained by generating events at parton level at LO and NLO accuracy and then matching 
 to {\sc Pythia8}. Uncertainties corresponding to PDF and scale variations are shown in the lower inset. }
 \end{figure}

\section{Conclusions}

The observation of multiple Higgs production at hadron colliders is a very challenging 
task, yet a crucial one to obtain key information on the form of the Higgs potential.
Rates for these processes are rather low and the accessible signatures swamped by large backgrounds.
In any case, any effort to gather information from measurements or bounds on these processes requires
accurate predictions for the SM total cross sections. In addition, differential distributions are needed not only to calculate
acceptances but also to improve the potential of disentangling these processes from the various backgrounds by selecting
the most sensitive regions in phase space. As in single Higgs production, the largest rates for multiple
Higgs production come from gluon-gluon fusion mediated by a top-quark loop. Loop-induced
processes pose one with the difficulty of obtaining higher order predictions, as these require the
computation of multi--loop Feynman diagrams. 

In this work we have first focused on Higgs pair production by considering different approximations 
 to improve the HEFT NLO calculation upon the infinite top-quark mass limit. We have introduced 
a reweighting procedure that allows the inclusion of the top-quark mass and width effects exactly. We have then applied it to $HH$ production
using the available information, {\it i.e.}  the exact (Born and real) one-loop amplitudes and the approximated two-loop matrix elements
to appropriately reweight events generated automatically by means of the MC$@$NLO method in the effective field theory. As a first result and
at variance with single Higgs production, we have found that including a non-zero top-quark width reduces the cross section by a couple of percents, the 
largest effect reaching -4\% just above the $t\bar t$ threshold. 

We have then performed a study to assess the relevance of various corrections and the accuracy of other approaches used in the literature to approximate NLO results in the FT. 
In particular we have compared to a Born reweighting, where only the exact Born results are used to improve upon the HEFT results.  At the total cross section level our best estimate
  gives a result about 10\% smaller than the Born-improved HEFT. 
We have then considered the difference between the two approaches for differential distributions, and found that including the exact real
emission matrix elements provides a better description of the phase space regions where hard emissions take place. We have then  
argued that total rates are improved too: by using an estimated form of the unknown virtual corrections in the FT using available results from single Higgs production, we have shown that our results are rather stable under variations of the unknown finite terms.  
Even though the effect of the missing two-loop virtual corrections on the total cross section 
cannot be quantified until these become available, comparing the different approximations allows one to conservatively associate an uncertainty of order 10\% 
with the calculation due to the missing top-quark mass effects. Note that this uncertainty should be 
quoted along with others until the exact NLO result becomes available.  Finally, we have applied our method to triple Higgs production, 
providing for the first time predictions for hadron colliders at NLO (+PS) accuracy in the SM. We have found a reduction of the theoretical uncertainties 
 and enhancements of the cross section similar to those of $HH$ production over a large range of collision energies.

\section*{Acknowledgements}
We are grateful to Claude Duhr, Giampiero Passarino, and Michael Spira  for stimulating discussions, and to Marius Wiesemann, Tobias Neumann and Michael Rauch for  their collaboration. We thank Stefano Frixione for many discussions and for comments on the manuscript.  This work has been supported in part by the Research Executive Agency (REA) of the European Union under the Grant Agreement number PITN-GA-2010-264564 (LHCPhenoNet) and PITN-GA-2012-315877 (MCNet).   
The work of MZ is partially supported by the ILP LABEX (ANR- 10-LABX-63), in turn supported by French state funds managed by the ANR within the ``Investissements d'Avenir'' programme under reference ANR-11-IDEX-0004-02.

\bibliographystyle{JHEP}
\bibliography{hhbib}
\end{document}